# Quantum Systems Engineering: A structured approach to accelerating the development of a quantum technology industry

M.J. Everitt[*], Michael J de C Henshaw and Vincent M Dwyer
*Quantum Systems Engineering Research Group, Department of Physics and The Wolfson School,
Tel: (+44) 01509 223325, e-mail: M.J.Everitt@physics.org*

**ABSTRACT**
The exciting possibilities in the field of new quantum technologies extend far beyond the well-reported application of quantum computing. Precision timing, gravity sensors and imagers, cryptography, navigation, metrology, energy harvesting and recovery, biomedical sensors and imagers, and real-time optimisers all indicate the potential for quantum technologies to provide the basis of a technological revolution. From the field of Systems Engineering emerges a focused strategy for the development cycle, enabling the existence of hugely complex products. It is through the adoption of systems thinking that the semiconductor industry has achieved massive industrial and economic impact. Quantum technologies rely on delicate, non-local and/or entangled degrees of freedom — leading to great potential, but also posing new challenges to the development of products and industries. We discuss some of the challenges and opportunities regarding the implementation of Systems Engineering and systems thinking into the quantum technologies space.

## 1. INTRODUCTION

The transistor, developed in 1947, was intended to replace valves, which were cumbersome, power hungry components essential to the operation of radios and other devices. The impact of such a simple change in technology was a revolution; on a daily basis almost everyone now makes use of the transistor, with communications, power production and health services all being reliant on these technologies. The emerging quantum technologies promise another such revolutionary change. There now exists a very real possibility that within the next few years we shall have new technologies, based on quantum physics, with incredible properties. Within state of the art laboratories it is possible to fabricate quantum devices capable of controlling nature at a fundamental level. Already, the UK's National Physics Laboratory has recently announced a quantum technology that provides a certified time distribution service for the financial sector eliminating the need for GPS. It is envisioned that consumer products will be developed resulting in e.g.: hyper-sensitive life-detectors; efficient discovery of new drugs; GPS that will work under water; clocks, with 100 times the accuracy of the atomic clock, available in compact portable devices. Ultra-high performance computing could be available on the desk top using low-power computing platforms. Internet fraud could be significantly reduced through digital signatures that cannot be cracked. New MRI-like scanners will give doctors a far more accurate picture of what is happening within the body leading to faster and more accurate diagnosis. Applications such as these clearly demonstrate the potential for quantum technologies, however much work needs to be done to manage the transition from state-of-the-art laboratory science into usable technology. This process can be greatly accelerated through the application of a new design approach that we term Quantum Systems Engineering. In this document we expand on this idea, and seek to explain the benefits of Systems Engineering to quantum technologies and why existing Systems Engineering methods will require tailoring in previously unexplored ways.

## 2. QUANTUM PHYSICS AND THE SYSTEMS ENGINEERING PROCESS

The Systems Engineering process is applied, as standard, to the development of the vast majority of complex products. It is an approach that takes into account factors such as requirements, functional analysis (this is the analysis of function – not the mathematical discipline!) and design synthesis (e.g. the bringing together of hardware and software). A *systems analysis and control* process is applied iteratively and recursively to manage the development of the "product". Systems Engineering has become so fundamental to product development that the US Department of Defence insist on *"a disciplined Systems Engineering process be used to translate operational needs and/or requirements into a system solution"* [1], and that *"The PM* [Project Manager] *implement a sound Systems Engineering approach to translate approved operational needs and requirements into operationally suitable blocks of systems. The approach shall consist of a top-down, iterative process of requirements analysis, functional analysis and allocation, design synthesis and verification, and system analysis and control. Systems engineering shall permeate design, manufacturing, T & E* [Transport and Environment]*, and support of the product. Systems Engineering principles shall influence the balance between performance, risk, cost, and schedule."* [2]. Classical Systems Engineering can, in part, be seen as the engineering of emergent phenomenon of a complex system, and it will be no different for its quantum equivalent. As the systems process continues, the design of interfaces between sub-systems (components) becomes crucial. This is combined with a "design for" mentality where detailed consideration is given to test, verification, validation, reliability, manufacture, end of life cycle and life cycle costing, support, risk and the impact of requirement changes, cost, etc. The the cradle-to-grave outlook of Systems Engineering considers the complete system lifecycle from the outset and takes into account high value components for recycling and reuse. In quantum technologies this is likely to be of significant concern. This approach has implications that are not usually considered by scientists working in the field but are accepted norms in many industries, and essential for achieving commercial benefit in the delivery of complex systems to market. Although the top-level system processes are applicable to all complex products [3], tailoring is always required to meet the challenges of specific product features.

To understand the challenge of this new tailoring, we need to consider a fundamental aspect of the quantum world known as entanglement. Any quantum systems can be described by operators in, and elements of, a vector space (strictly speaking a Hilbert Space). Composite quantum systems (which may include the environment and device control circuitry) are formed through tensor products of these spaces. Two of the important consequences of this are (a) that the number of degrees of freedom scales exponentially fast and (b) that composite quantum systems cannot be considered as a collection of isolated entities, as quantum states may not be representable as the product of states of the individual subsystems; these states are termed 'entangled'. While (a) leads to the massive processing power expected of quantum computing algorithms, it concomitantly leads to challenges in modelling and simulation which are key components of the Systems Engineering process. More fundamentally important however is (b), as entanglement is in direct conflict with the need to specify clear system boundaries, their interfaces and requirements. We will refer to these two issues as the *modelling* and *interface problems* respectively. As discussed later, we see the possibility for specific Quantum Systems Engineering strategies in, for example, design for reliability or test, but it is the difficulty of being able to clearly specify a systems boundary that presents one of the greatest challenges to developing a Systems Engineering approach to any but the simplest of quantum technologies.

Traditional systems engineering was concerned with single systems (a product perspective) in which the system boundary, referred to as the System Of Interest (SOI). But in recent times, as the world becomes more highly interconnected, the problem of defining the SOI becomes multi-faceted for Systems of Systems [4]. Esoteric interactions between entangled quantum systems creates a new level of complexity in defining the SOI. In the next section we discus a few example topics that will form part of Quantum Systems Engineering in more detail. It is worth noting that these areas are by no means exhaustive. There is another side to Systems Engineering that is not covered in this document and that is the community aspect as exemplified by the International Technology Roadmap for Semiconductors (ITRS). While various initiatives exist, as a community, we lack the coordinated effort of the ITRS.

### 3. SOME ELEMENTS OF QUANTUM SYSTEMS ENGINEERING

### 3.1 Abstraction & High Level Design

Quantum mechanics remains an abstract and conceptually challenging subject, despite its mathematical description being relatively straightforward. Indeed no consensus has yet been agreed about the ontological interpretation of the subject. This coupled with an incomplete understanding of issues such as the measurement problem and characterising entanglement within many particle systems, will act as a barrier to entry for those wishing to work in quantum technologies but without relevant background. That said we are aware of one notable example where the very abstract concept of topological quantum computation has been encoded within a computer game enabling those with no understanding of the underlying physics to contribute to the optimisation of advanced algorithms (http://www.mequanics.com.au). In order to enable high level design strategies to be developed we will need to produce tools capable of taking advantage of similar user interface design.

### 3.2 [Design for] Reliability

Reliability engineering is used to ensure products work to specification in terms of accuracy, working lifetime, chance of failure, etc. Being able to design quantum systems for reliability will involve understanding all the various possibilities of a device failing. It will require, for example, understanding of: all failure modes and what drives them (physics of failure models); all manufacturing tolerance and physical purity implications; the probabilities of mode occurrence and physical parameters that govern them; any mechanisms which might accelerate failure such as increased temperature or stress and strain; the cumulative distribution function (cdf) of failure times for a given set of acceleration parameters and a means of deceleration back to use conditions. One of the salient features of the quantum state is its fragility. The presence of environmental degrees of freedom, such as environmental and circuit noise, are an acute problem especially as quantum systems are likely to grow in complexity. The distributed nature of entangled quantum information means that unexpected failure modes are likely to be a problem. A good analogy of where Systems Engineering will be of use can be found in designing components to minimise the damaging effects of Electrostatic Discharge (ESD). Here it was found that optimisation of an individual unit against ESD leads to less than optimal system designs [5]. It is evident that understanding the operation of the system as a whole, including deployment, is of central importance. Quantum technologies will be substantially more prone to failure but design strategies such as these, adapted to take into account the entangled nature of the quantum state, will be essential. Furthermore, problems will arise when integrating coupled quantum systems and their associated control and read-out circuitry. That is, the environment of a quantum technology includes the associated circuitry, packaging, off the shelf power supplies together with other influences of the outside world. In analogy with protecting circuits against electrostatic discharge, it is likely that the optimal design for a component included in a production device will be different from that of units tested under laboratory conditions. So that, for complex quantum-technology systems, there will be emergent phenomena requiring new reliability engineering design paradigms – where, as an example engineered environments such as in [6,7] could lead to novel strategies to protect against some failure modes.

### 3.3 [Design for] Test and Certification

Even setting aside the subtleties that exist within the measurement theory of quantum objects there will be practical difficulties in realising test and certificated protocols for chosen technologies. There exists a couple of theoretical studies that have begun to investigate this problem using quantum statistics as a basis for certification [8, 9]. Such approaches may well find utility for those systems who statistics we understand from either theoretical or classical numerical modelling.  It will be in areas where such an analysis is not possible, such as quantum simulation for drug discovery or arrays of quantum sensors, that test and certification becomes particularly challenging. To be more explicit if we consider a quantum emulator of molecular dynamics for a system that exceeds our classical computational power, how do we develop figures of merit that enable us to make value judgements on the output of our simulation? In an additional step, it will also be important to include general design principles which allow access to debug and fault identification which classical require the notion of system state and its replication where the quantum no-cloning theorem may limit existing methods.

### 3.4 [Design for] Scalability and the Interface Problem

The emergent properties of large quantum and hybrid quantum-classical systems are still not well understood. Any product will be a composite quantum-classical system incorporating measurement, feedback and control and will need to be deployable in a range of scenarios. Theoretically, we have increasingly sophisticated models of both closed and open systems that can incorporate feedback and control (including back-action). These models have great predictive power for small quantum systems, but become rapidly ineffective as systems increase in scale. Quantum systems lack a set of hierarchal models, from components to systems, that can be used to accelerate the design process. In order to be able to develop such a hierarchy, it is necessary to identify system boundaries and capture the interaction between components at differing levels of complexity. As has been discussed above, the interface problem makes this uniquely challenging as components can not always be separated from the rest of the system and the systems state will not be localisable into states for each unit.

### 3.5 Modelling and Simulation

A key to the rapid design of any advanced technology is the use of Computer Aided Engineering. Within quantum technologies, as with VLSI chip design, this become particularly important in removing from the quantum engineer the need for a sophisticated skill-set for modelling open quantum systems. While much of the necessary physics is well developed, application to collections of quantum systems and their classical feedback and control circuits remains a grand challenge (extending beyond the well-known quantum simulation problem to include quantum measurement). It is essential for the community to have access to other computational methods and more effort is needed to generate general design tools rather than bespoke problem solvers.

### 3.6 Requirements Specification for Quantum Devices

Within the semiconducting device industry, the military, and increasingly in other disciplines, the first stage of product design is captured by a set of customer (or design) requirements which are then developed into a Systems Requirement Specification (SyRS). Well-formed high-level (customer) requirement statements will not mention quantum or any other technological aspect of the solution, these emerge as the details are developed into the SyRS. The document takes the form of a *"...collection of requirements that constitutes the specification and its representation and acts as the bridge between the two groups and must be understandable by both the customer and the technical community. One of the most difficult tasks in the creation of a system is that of communicating to all of the subgroups within both groups, especially in one document. This type of communication generally requires different formalisms and languages"* [10]. The SyRS must capture all aspects of the design process including: process inputs and outputs; requirements analysis; functional analysis and allocation; design synthesis and verification/validation. In formulating a complete requirements analysis consideration needs to be given to specific aspects related to e.g. design, implementation, disposal and recycling, and maintenance, all of which will affect the SyRS by imposing constraints. Quantum physics will affect the requirements analysis at the stage of creating the SyRS in a non-trivial way. It is apparent that the requirement to maintain quantum coherence will impact the performance and design requirements. However, there will also be derived requirements that act in both directions of the design process (e.g. [top-down] the device must be usable by a single operative in the field or

[bottom-up] limits imposed on the maximum ambient operating temperature). Given the subtleties of quantum physics, consideration needs to be given to how to generate a well formulated SyRS (where each requirement is achievable, clear, independent of physical realisation, testable and verifiable, consistent and appropriate). The challenge here is exemplified by the fact that the specification of interfaces will need to take into account the sensitive nature of quantum states. Here environmental effects, the physics coupling quantum devices to each other and the system's classical circuitry, must be taken into account. There is also the possibility of other effects such as parasitic entanglement arising within the system. Such considerations are needed to develop a SyRS capable of capturing all aspects of the design process and thus be fit for purpose. Regardless of these issues, as quantum technologies still mainly reside in the research sphere, the move to using a SysRS that would refocus the mind of quantum technologist from prototype to product, would be of value in reducing the occurrence of designing prototype solutions that are not well suited to production and would require re-engineering.

### 3.7 Ontology

With terms as apparently simple as position, angular momentum, and spin having very different meanings within quantum and classical physics, and terms such as functional analysis meaning different things in mathematics and engineering, the potential for confusion for those collaborating across disciplines is high. The realisation of quantum devices will necessarily require engineering to address phenomena that have not been considered hitherto: effective communication between science and engineering will be crucial and, to this end, the development of an ontology for QSE, that will enable precision in the definition of (for instance) systems requirements, is an essential step in the elaboration of QSE.

## CONCLUDING REMARKS

Quantum 2.0 Technologies, by definition, leverage quantum superposition in a way that is fundamental to their operation. An understanding of how such physics will impact the emergent properties of complex quantum devices has yet to be developed. Because of the special requirement of maintaining a necessary level of quantum coherence, readout and control, current approaches to the tailoring of Systems Engineering will not encompass all the methods needed to produce devices dependent on such physics, and a new Quantum Systems Engineering framework is needed that, due to the quantum interface problem, will be very different in some aspects from existing formalisms. Moreover, we must remove from this framework the need for a deep understanding of the underpinning theoretical physics in high level design processes as this is key to removing barriers to entry for the industry. Adopting a Systems Engineering mind-set will, regardless of the challenges involved, provide the quantum engineer with a structured approach to accelerate the development of robust quantum solutions.

## ACKNOWLEDGEMENTS

We thank the School of Science and Wolfson School at Loughborough University for their generous support of the centre for doctoral training in Quantum Systems Engineering.